\documentclass[preprint2]{aastex}

\received{2002 August 13}
\begin{document}
\newcounter{bbb}

\newcommand{\ade}{{\mbox{\sc ademis}}}

\newcommand{\map}{{\mbox{\sc mappings~i}c}}
\newcommand{\mapI}{{\mbox{\sc mappings~i}}}
\newcommand{\is}{{\mbox{\sc is_galaxev}}}

\newcommand{\etal}{et al.\ }
\newcommand{\msun}{\hbox{M$_{\odot}$}}
\newcommand{\lsun}{\hbox{L$_{\odot}$}}
\newcommand{\zsun}{\hbox{$Z_{\odot}$}}
\newcommand{\ia}{\'{\i}}
\newcommand{\hb}{\hbox{H$\beta$}}
\newcommand{\ha}{\hbox{H$\alpha$}}
\newcommand{\msup}{\hbox{$m_{up}$}}
\newcommand{\minf}{\hbox{$m_{inf}$}}
\newcommand{\mlow}{\hbox{$m_{l}$}}
\newcommand{\oii}{\hbox{[\ion{O}{2}]}}
\newcommand{\oiii}{\hbox{[\ion{O}{3}]}}
\newcommand{\nii}{\hbox{[\ion{N}{2}]}}
\newcommand{\sii}{\hbox{[\ion{S}{2}]}}
\newcommand{\hnd}{\hbox{\ha$+$\nii}}
\newcommand{\HII}{\hbox{\ion{H}{2}}}
\newcommand{\hii}{\hbox{\HII}}

\newcommand{\wl}{\hbox{$W_{\lambda}$}}
\newcommand{\joo}{JFFC}
\newcommand{\joof}{Jansen et\,al.\,(2000a,b)}
\newcommand{\jooa}{Jansen et\,al.\,(2000a)}
\newcommand{\joob}{Jansen et\,al.\,(2000b)}

\shortauthors{Magris, Binette \& Bruzual}

\shorttitle{\ade: A library of emission-line models}  

\title{ADEMIS: A Library of Evolutionary Models for Emission-Line
Galaxies: I- Dustfree Models }

\author{Gladis Magris C.}
\affil{Centro de Investigaciones de Astronom{\'\i}a (CIDA),\\
Apartado Postal 264, M\'erida 5101-A, Venezuela}
\email{magris@cida.ve}

\author{Luc Binette}
\affil{Instituto de Astronom\ia a, Universidad Nacional Aut\'onoma
de M\'exico, \\ Apartado Postal 70-264, 04510 D. F., M\'exico,
M\'exico}
\email{binette@astroscu.unam.mx}

\and

\author{Gustavo Bruzual A.}
\affil{Centro de Investigaciones de Astronom{\'\i}a (CIDA),\\
Apartado Postal 264, M\'erida 5101-A, Venezuela}
\email{bruzual@cida.ve}

\begin{abstract}

We present an extensive set of stellar population synthesis models,
which  self-consistently include the (optical--far-UV) continuum
emission from stars as well as the resulting emission-line spectrum
from photoionized gas surrounding massive stars during their main-
sequence-life time. The models are presented as  a compiled
library, \ade, available electronically.  \ade\  contains the
equivalent widths and the intensities of the lines
\oii\ $\lambda\lambda$3727, \hb, \oiii\ $\lambda$5007, \ha, and 
\nii\ $\lambda$6584, which were calculated assuming a metallicity of
0.2, 0.4, 1.0, and 1.5 \zsun\ and a wide range of ionization
parameters. We investigate the regime of continuous star formation,
assuming a Salpeter initial mass function, whose upper mass limit is
 an input parameter.  The calculated equivalent width
models, which are function of the burst age, are compared with the
Jansen et al. atlas of integrated spectra for nearby galaxies.  We
reproduce the observed properties of galaxies along the full Hubble
sequence and suggest how the metallicity and age of such galaxies
might be roughly estimated.

\end{abstract}


\section{Introduction}

The presence of emission lines is a conspicuous feature of optical
spectra of any galaxy that is currently forming stars. The lines
result from the reprocessing of the UV ionizing radiation from the
most massive of the young stars.  Therefore, emission lines carry
imprints of both phases, the newly formed stars versus the `excited'
interstellar medium (ISM), in a single observable quantity, the
integrated optical spectrum. It is for this reason that the modeling
of the emission-line spectra occurring in star-forming galaxies must
include the interaction between both phases (stars $+$ gas) in a
self-consistent manner. Partly for this reason, this paper will
emphasize the study of equivalent widths\footnote{Equivalent widths
present the advantage of relating any line flux with an observable
quantity at the {\it same} wavelength, namely the underlying
continuum. There are quite a few papers in the literature that are
concerned with the calibration of the \oii\ equivalent width and whether
it can
be used as a reliable tracer of star formation. Such diagnostics of
\wl(\oii) in the UV would be particularly valuable at redshifts where
\hb\ or \ha\ become inaccessible.} (EWs or \wl). The
optical stellar continuum underlying the emission lines consists of
much more than just the stars that ionize the gas. Actually, in any
evolved stellar population, the optical and red continua are dominated
by red giants and low-mass main sequence stars. In order to model the
observed continuum properly, one must therefore consider the past
stellar formation history of the whole stellar population, an
essential step if the emission-line equivalent widths are to be
calculated reliably.

Working toward these goals, Garc\ia a Vargas, Bressan
\& D\ia az (1995), Stasi\'nska \& Leitherer 1996, Stasi\'nska,
Schaerer \& Leitherer (2001) built models in which the spectral
energy distributions (SEDs) generated by stellar evolutionary
synthesis codes were used as input to photoionization calculations of
emission-line strengths.  These models aimed mainly at constraining
the ages and metallicities of isolated extragalactic \hii\ regions
that are characterized by the absence or weakness of an old stellar
population, suggesting a present-day strongly enhanced star formation rate.  In
short, these earlier models followed the evolution of spectral
properties of an instantaneous burst of star formation. Recently,
Kewley \etal (2001), Zackrisson \etal (2001) and Moy, Rocca-Volmerange
\& Fioc (2001) introduced models in which the star formation rate   is
constant for a few megayears, without interruption. These were used to derive
the properties of starburst galaxies.

In an attempt to include the forgotten dust phase, Magris (1993)
combined the spectral evolution synthesis code of Bruzual \& Charlot
(1993) with the photoionization code \mapI\ (Binette, Dopita \& Tuohy
1985). The particularity of these models is that they included the
effects of dust on the line- and continuum-emission in a
self-consistent manner.  On the other hand, only the solar metallicity
case was studied, since detailed non-solar metallicity tracks and
stellar spectral libraries were not complete at that time.

Barbaro \& Poggianti (1997) have presented a model that is used to
calculate the emission lines of \oii\ $\lambda\lambda$ 3727 and of the
Balmer series along with photometric quantities. The purpose was to
estimate the present day star formation rate in late type galaxies. In
their model, the star formation rate and the chemical evolution of the
stellar phase for each Hubble type have been derived from their own
chemical evolution model, assuming, on the other hand, a fixed solar
composition for the ionized gas.

Charlot \& Longhetti (2001) computed the line and continuum emission
from star-forming galaxies and presented a family of estimators of the
star formation rate, the nebular oxygen abundance, and the optical
depth of the dust, in galaxies, as a function of the available
spectral information.  Their model is generally successful in
reproducing the observed properties\footnote{See Charlot \etal (2002)
for a more recent application of this model to the determination of
the star formation rate.} of template galaxies of different
types. Their analysis, however, did not include the modeling of
equivalent widths that are particularly suitable for analyzing the
interrelated properties of the stellar and nebular phases.

We have developed a model along similar lines, which calculates, for an
arbitrary star formation rate, both the nebular emission-line spectrum
of the \hii\ regions associated to the newly born massive stars and
the SED of the underlying stellar population.  However, rather than
discussing line ratios and stellar energy distributions separately,
our work will emphasize the relationship between the two phenomena, by
presenting our model results as a library, \ade, of emission-line
equivalent widths. As supplementary
information, the tables also include
the line luminosities of the most
prominent species observed in normal star-forming nearby galaxies,
i.e. \oii\ $\lambda\lambda$3727, \hb, \oiii\ $\lambda$5007, \ha, and \nii\
$\lambda$6584. The models are characterized by a metallicity range of
0.2, 0.4, 1, and 1.5 times solar (gas $+$ star) and consider rates of
star formation that are constrained by both the line and the stellar
continuum emission, of galaxies of all Hubble types.

The method used to derive our evolutionary models is presented in \S
2.  A description of the results from \ade\ follows in \S 3
accompanied by a comparison of the models with the Jansen et al. (2000a,b;
hereafter \joo) 
atlas of nearby galaxies. The conclusions are summarized in \S 4. A
more extensive version than the adjoining tables of  \ade\
models is available on the Web.

\section{Computational Method}

The method used to derive the nebular synthetic spectrum consists
first in obtaining the ionizing spectrum corresponding to the hot main
-sequence stars of an arbitrary stellar population.  This ionizing
spectrum is subsequently used by a photoionization code to compute the
emission-line spectrum. The line fluxes are then combined with the
integrated continuum spectrum of the stellar component to provide us
with the equivalent widths of the emission lines of interest.

\subsection{The Population Synthesis Code}

One of our aims is to  derive a nebular spectrum fully consistent 
with the young stellar population from which arise 
\HII\ regions within an evolved normal galaxy. Special care must
therefore be given to the population synthesis of the star-forming
regions. For this purpose, we use the latest version of the
evolutionary population synthesis code of Bruzual \& Charlot (2003,
hereafter BC03).

With this code, we can derive the spectrophotometric properties of any
specified stellar population. For a given star formation rate (SFR)
and initial mass function (IMF), a model consists of a set of spectra
of the integrated stellar population calculated at 221 different ages
$t$. Each such spectrum is defined at a common set of wavelength
points covering the range 5--100 $\mu$m.  The BC03 code,
furthermore, can store   separately the  contribution of stars in different evolutionary stages. In the current
work, it was necessary to generate an integrated spectrum for all the
stars and another one for the subpopulation of main-sequence (MS)
stars only. This latter spectrum was  essential to derive the ionizing
SED, which excites the \HII\ regions and whose contributors consist
exclusively of massive MS stars. This procedure excludes, for
instance, other types of hot stars that generate ionizing photons,
such as planetary nebulae nuclei, white dwarfs, or core helium-burning
stars, which are irrelevant to young \HII\ regions.  In a separate
paper, we have verified that the small \ha\ widths of normal
ellipticals might be explained by photoionization from post-AGB
stars alone (Binette et al. 1994).

The time-evolution of the stellar population is followed using the
Girardi \etal (2000) stellar tracks for low- and intermediate-mass
stars (i.e. from 0.15 to 7 \msun), which are the most recent set from
the Padova group, and the Bertelli \etal (1994) tracks for stars from
9 to 120 \msun. The advantage of these models with respect to earlier
ones is that the data-base is computed with a homogenous input
physics for all stellar tracks in the grid. For instance, moderate
convective overshooting from convective cores and mass loss from
massive stars are taken into account.  In the most recent version used
here, these tracks have been improved further with the adoption of new
low-temperature opacities (see Girardi \etal 2000 for details)

The Kurucz stellar atmospheres spectra library (distributed by
Lejeune, Cuisinier, \& Buser 1997) has been adopted to compute the
stellar properties of stars with $T_{eff} < 50,000$ K. For the higher
temperature stars, in order to give an adequate treatment to the
ionizing continuum of hot MS stars, we use the Rauch (1977) NLTE model
atmospheres\footnote{Also available at 
\url{http://astro.uni-tuebingen.de/$^\sim$rauch}/. } for $\lambda <
2000$ \AA, and a simple black body spectrum for larger wavelengths. We
recognize that large differences exist in the ionizing flux between
LTE model atmospheres that include only H, He, C, N, O, and Ne, (like
the Hummer \& Mihalas (1970) models used in the emission-line
synthesis of Magris (1993)) and the NLTE Rauch (1997) models, which
include the opacity of heavier elements from Li to Ca. Despite these
differences, we find, however, that the general properties of the
ionizing flux that emerges from MS stars in a composite
stellar population are not significantly affected. The reason is that
high-mass stars have a relatively short lifetime for $T_{eff}$
hotter than $50,000$ K, and their contribution to the integrated
spectrum of a composite stellar population becomes insignificant for any
lasting star formation burst.

\subsection{The Photoionization Code}

To compute the nebular spectrum, we use the multipurpose
shock/photoionization code \map\ described by Ferruit et al. (1997); 
see also Binette et al. 1985), in which the ionization
and thermal structure of the nebula are computed in a standard fashion,
assuming local ionization and thermal equilibria throughout the
nebula. The calculation is made at a progression of depths
in the nebula, until all the ionization radiation has been
absorbed.

For the nebular gas distribution, we adopt a simple slab geometry as
in the work of Shields (1974), who argued that the integrated line
spectrum was not significantly different from that calculated with the
more usual spherical geometry.  It is also the geometry preferred for
the Orion nebula as modeled by Baldwin et~al. (1991) or Binette, González-Gómez \& Mayya (2002). The rationale is that \HII\ regions are rather chaotic in
appearance, showing evidence of strong density fluctuations with much
of the emission coming from condensations or ionized ``skins'' of
condensations.  The relation between the dimensionless ionization
parameter $U$, as defined in this work\footnote{We define the
ionization parameter as follows, $U= {Q_{H^0}}/({4 \pi r^2 c \, n_H})$, where
$Q_{H^0}$ is the photon luminosity (quanta s$^{-1}$) of the stellar 
cluster, $c$ is 
the speed of light, and $r$ is the distance separating the photoionized gas
of density $n_H$ and the ionizing stellar cluster.}, and the emission
lines, is also more straightforward and intuitive in the plane-parallel
slab case.

Our set of solar abundances is taken from Anders \& Grevesse
(1989). When varying metallicity, we will assume that the abundances
scale with metallicity in a linear fashion for all elements except
helium and nitrogen. The He/H ratio will be kept constant at the solar
value.  The observed behavior of the N/O ratio in local and external
\hii\ regions suggests that the nitrogen abundance increases via
secondary nucleosynthesis for metallicities above 0.23 solar (van Zee,
Salzer \& Haynes 1998; see Henry and Worthey 1999 for a review). To
represent such N/O behavior, we adopt the following relations (Dopita
\etal 2000):

$$\log ({\rm N/H}) = - 4.57 + \log (Z_{gas}/Z_{\odot});~~~\log (Z_{gas}/Z_{\odot}) \leq -0.63$$
$$\log ({\rm N/H}) = - 3.94 + 2 \log (Z_{gas}/Z_{\odot});~~~{\rm otherwise}$$
\noindent where $Z_{gas}$ denotes the gas phase metallicity.

\subsection{The Jansen et al. atlas  as comparison sample}

We will compare the predictions of our model with the
spectrophotometric atlas of nearby galaxies of \joo, which is ideal
for this purpose, since it includes a representative sample of
galaxies spanning the whole Hubble sequence and extending to lower 
luminosity systems than the Kennicutt (1992) sample. This sample 
is frequently referred to as the Nearby Field Galaxy Survey (NFGS).
Such a
comparison with the \joo\ data will allow us to test the global
properties of our models in the observational framework of a wide
range of galaxy luminosities.  The full sample includes 196 nearby
galaxies from E to Irr, of luminosities in the range $M_B = -14$ to
$-22$.  However, of the 141 objects with detected \ha\ emission, we
will consider only those galaxies for which the equivalent widths
EW$(\ha) > 10 $ \AA\ and EW$(\hb) > 5 $ \AA. This will ensure a
reliable measurement of Balmer decrements according to Jansen, Franx,
\& Fabricant (2001) and leave us with a sample of 85 objects. The
atlas contains the equivalent widths and line ratios with respect to
\hb\ of the following lines: \oii\ $\lambda\lambda$ 3727, H$\delta$,
\hb, \oiii\ $\lambda$4959, \oiii\ $\lambda$5007, He~{\sc I}\
$\lambda$5876, {\char91 O~{\sc I}\char93}\ $\lambda$6300, \nii$\
\lambda$6548, \ha, \nii$\ \lambda$6584, 
\sii$\ \lambda$6716, and \sii$\ \lambda$6731.

Since Jansen et al. (2001) have reported that the Balmer
emission-line fluxes and equivalent widths given in \joo\ were only
partially corrected for stellar absorption, we have applied the
additional correction suggested by these authors of 1 and
1.5 \AA\ (EW) for \hb\ and \ha, respectively.

We corrected all line ratios for interstellar reddening by comparing
the observed Balmer decrement, \ha/\hb, with the case B recombination
value of 2.85 (appropriate for $n_e \approx 10^2 - 10^4 cm ^{-3}$ and
$T = 10^4 $K; Osterbrock 1989), and using the Nandy et al. (1975) Galactic
extinction curve as tabulated by Seaton (1979).

\subsection{Grid of photoionization calculations}

Photoionization models were calculated using the various ionizing SEDs
computed by the BC03 code. These SEDs differ by the age $t$ of their
stellar population and correspond to MS stars, which were all formed at
$t =0$, following a simple Salpeter (1955) IMF that covers a mass range
comprised between a lower and upper mass limit, $m_{low}$ and
$m_{up}$, respectively. The adopted lower limit is $m_{low}= 0.09$
\msun\ in all models.  
Because of the crucial role played by the upper mass limit on
the hardness of the energy distribution, we have explored a wide range
of upper mass limits: $m_{up} =$ 120, 100, 90, 80, 70, 60, 50, and 40
\msun.  As for the metallicity of the stellar population, 
$Z_{\ast}$, we have calculated solar metallicity models
($Z_{\ast}=0.019$) as well as metal-poor ($Z_{\ast}=0.008$ and 0.004) and
metal-rich ($Z_{\ast}=0.03$) models.

\subsubsection{Usefullness of the hardness parameter $\eta$}

We measure the hardness of the ionizing spectrum using the ratio between the
flux of stellar photons (quanta s$^{-1}$) emitted above 24.6 eV and
those emitted above 13.6 eV, $\eta = Q_{He^0}/Q_{H^0}$.  The hardness
$\eta$ can be associated with the cluster {\it equivalent effective
temperature}, frequently used by \HII\ region modelers (e.g. Shields
\& Tinsley 1976; Stasi\'nska 1980; Campbell 1988).  

In Fig. 1a, we show the hardness, $\eta$, as a function of \msup, of
various ionizing distributions that are later used in photoionization
calculations. An instantaneous burst with a Salpeter IMF was assumed.
Each curve corresponds to a different stellar track metallicity,
$Z_{\ast}$, as labeled.  The different curves show a reduction in
$\eta$ with increasing metallicity, because the higher $Z_{\ast}$, the
cooler the stellar evolutionary tracks, and therefore, the softer the
SED (and the lower $\eta$).  This behavior has previously been
reported by McGaugh (1991), who used the zero-age point of the stellar
evolution tracks of Maeder (1990) to compute the ionizing radiation
field of a cluster of stars distributed as the Salpeter (1955) IMF.
Even though the details about our respective ionizing spectra may
differ, it is found in this work as in McGaugh, that for a given SED
with a certain $\eta$, it is possible to encounter another SED with
the same $\eta$ if we decrease (or increase) appropriately {\it both}
\msup\ and $Z_{\ast}$.  This interesting property, inferred from
instantaneous burst models, tells us that both parameters are in fact
degenerate.

Figure 1b shows the {\it temporal} evolution of $\eta$ for a model
with a continuous SFR calculated with a Salpeter (1955) IMF, $\msup\ =
125 \msun$, and the metallicities as labeled.  We can see that $\eta$
reaches a constant asymptotic value after $\approx 10^7$ \, yr, due to
the equilibrium between the number of new born massive stars and the
number of them evolving off the MS.  The same kind of evolution occurs
with any star formation rate provided it is continuous and either
constant with time or monotonically decreasing.  In Fig. 1b, we also
show the behavior of $\eta(t)$ for the case of a solar metallicity
instantaneous burst, for comparison (dashed line).

In our extensive grid of photoionization calculations, $U$ was
varied over the wide range of $-4 \le {\rm log ~} U \le -1.5$ although,
in practice, a more limited range suffices, as independently shown in
the modeling of \HII\ regions by different authors (e.g.,
Evans \& Dopita 1985; Campbell 1988).  

As for the nebular gas metallicity, $Z_{gas}$, as a rule we  adopt the same
value as for the stellar atmospheres, and  we will thereafter use $Z$
to refer to the metallicity of either component. 

As a result of these computations, we obtained an extensive grid of
emission-line spectra parameterized according to $U$, $\eta$, and
$Z$. We verified that it was sufficient to read the appropriate
emission line fluxes from our grid of photoionization models,
using $\eta$ as the lookup value.  Models at intermediate values
of the parameters $U$, $\eta$, and $Z$ are obtained by simple
interpolation (in the log-plane) within the grid points.
This procedure allowed us to do away
with calculating separate new photoionization calculations each time
$\eta$ changed.

\subsubsection{Two standard diagnostic diagrams}

In Figure 2 we show a standard diagnostic diagram of the relationship
between\\ \oiii$\lambda 5007 $/\hb\ and \oii$\lambda\lambda
3727$/\hb. The solid lines join discrete models (open circles)
calculated by using the \msup\ value that appears in panel b near the right
extremity of each line. Different \msup-values correspond to different
$\eta$-values with $\eta$\ strongly dependent as well on metallicity as
Figure 1 already showed.  The dotted lines join points of constant $U$
(indicated as log $U$-values near the upper end of the curves).  With
a range of ionization parameter of about a factor of 10 ($-3.25 <$ log $U
< -2.25$) as plotted in this figure, and a range of metallicity
between 0.4 and 1.5 \zsun, our photoionization calculations encompass
the dispersion in the
\oiii/\hb\ versus \oii/\hb\ relation observed in the
\joo\ sample of galaxies.

The same range in ionization parameter and metallicity characterizes
the extragalactic \HII\ region sequence of Dopita \etal (2000).  Similarly,
Bresolin, Kennicutt \& Garnett (1999), who used a different but scalable definition
of $U$, applicable to their geometry, concluded that a range of
only a factor of 10 in $U$ suffices to reproduce the optical emission-line 
spectrum of their sample of spiral galaxy \HII\ regions.
Our models also happen to overlay the correlation presented by Kobulnicky,
Kennicutt, \& Pizagno (1999) in their  \oiii/\hb\ versus \oii/\hb\
diagram, which superposes  indifferently individual \HII\ regions 
and galaxy integrated ratios. 

In Figure 3 we show the line ratio diagnostic diagram of
\oiii/\oii\ versus \nii/\oii, which   Dopita \etal (2000) suggests
as being optimal for separating variations in the ionization parameter
from variations in chemical abundance.  The upper panel has been
calculated with a simple zero-age stellar population, while the lower
panel used a $10^7$ y old population formed according to a continuous
SFR prescription. The \oiii/\oii\ ratio has been established to be
useful for determining the ionization parameter (e.g. McGaugh 1991),
while the \nii/\oii\ ratio was found to be virtually independent of
the ionization parameter, albeit very sensitive to the abundance for
metallicities above 0.23 solar.  This Figure indicates that the
galaxies in the \joo\ sample span a range in abundances from 0.2 solar
up to solar or possibly 1.5 times solar, without any evidence for much
higher metallicity objects\footnote{The alternative \nii/\ha\ ratio
has previously been suggested as an abundance estimator for low
metallicity galaxies by Storchi-Bergman, Calzetti \& Kinney (1994; 
see also van Zee et al. 1998; Raimann et al. 2000; Denicol\'o,
Terlevich \& Terlevich 2002).  However, for abundances greater than
0.5 solar, a ``fold'' in the behavior of \nii/\ha\ is clearly present
in the diagnostic plot of \oiii/\hb\ $vs.$ \nii/\ha, as revealed by
Dopita et al. (2000).}.  As in Figure 2, log~$U$ spans the range
$-3.25$ to $-2.25$.

Interestingly, the continuous SRF models of Fig. 3b overlap better the
observational points than the zero-age models of Fig. 3a. This simply
results from having a softer SED (smaller $\eta$) in the continuous
SFR case, which produces models with a lower \oiii/\oii\ ratio.

\subsubsection{The synthesis of line spectra, stellar spectra and EWs}

The SED of a given stellar population with age $t$ and arbitrary SFR,
IMF, and $Z$ is computed using the BC03 code.  From there, the SED
corresponding exclusively to MS ionizing stars is used to derive
$\eta$ and $Q_{H^0}$, the number of ionizing photons, which then
allows us to infer the nebular emission-line spectrum by simple
interpolation within the grid of \HII\ region models, as described in
\S 2.4.1. Finally, each emission-line equivalent width
$W_{{\lambda}_i}(t)$ of line $i$ at wavelength $\lambda_i$
corresponding to the above evolutionary age $t$ is derived by dividing
the line luminosity $L_i^l(t)$ by the SED luminosity $\bar
L^{\ast}(\lambda_i,t)$ of the integrated stellar population (i.e. all
stars included) averaged over a fixed narrow band-pass
$\Delta_{\lambda_i}$.


\subsubsection{The concept of  ``equivalent \HII\ region''}

In the present model, we are assuming that the integrated emission
line spectrum of galaxies is well represented by an ``equivalent \HII\
region,'' ionized by a cluster of MS stars with the same mass spectrum
as the whole galaxy.  In support of this assumption, it has been shown
by Kobulnicky et al. (1999) that the integrated \oii\
and \oiii\ line luminosity mimics that of individual
\hii\ regions, and that the chemical abundances of the gas phase 
derived from individual nebulae or from integrated galaxy spectra coincide
at least within the intrinsic uncertainties of the proposed abundance
determination method.  For ages older than 10 Myr and for a continuous
star formation rate, the number of newly born massive stars relative
to those having aged and producing less ionizing photons, achieve a
near steady state regime of constant hardness, as can be seen in the
right-hand panel of Figure 1, which shows the time evolution of
$\eta$ for different continuous SFR.  In this adopted scenario, the
shape of the ionizing spectrum depends therefore only on the IMF and the
metallicity of the stellar population.  
We are implicitly assuming that the error
introduced by calculating the nebular spectra using a unique (but  variable)
galaxy age $> 10$ Myr (rather than taking into account a
distribution of \hii\ region ages), is small in comparison with the
uncertainties introduced by the other nebular parameters. Actually,
our models are intended for comparison with normal galaxies, which are
massive enough to minimize the dispersion because of statistical
fluctuations in the sampled population.  However, as has been pointed
out by Cervi\~no \etal\ (2002), this question is susceptible to be
investigated in more detail.  

Interestingly, if one seeks the equivalent \msup\ of a zero-age
instantaneous burst (Fig. 1a), which would reproduce the same SED
hardness $\eta$ as that from the corresponding asymptotic value in the
continuous SFR case with $\msup = 125 \msun $ (Fig 1b) and {\it with
the same} $Z$, one finds that this zero-age equivalent \msup\ is
$\simeq 60 \msun $. This result was confirmed for all the
metallicities studied in this paper. It follows that the continuous
SFR models of Fig. 3b with $\msup = 125 \msun $ are, to all practical
purposes, equivalent to a zero-age instantaneous burst with $\msup
\simeq 60
\msun$.

\section{Model results and discussion}

We first discuss the principal results of our model predictions based
on exponentially declining SFRs and then proceed to compare these
with the spectrophotometric atlas of \joo.

\subsection{An atlas of calculated equivalent widths for non-instantaneous bursts}

Table 1 summarizes the contents of our model grid \ade\footnote{Also
available as an ASCII file via ftp to
\url{http://www.cida.ve/$^\sim$magris/ademis} .}. 
The models listed correspond to the Salpeter IMF, with $\msup = 125
\msun$, and a SFR $\Psi(t)$ exponentially declining with time:
\begin{equation}
\begin{array}{c}
\Psi(t)=\tau^{-1}{\rm exp}(-t/\tau) \; ,
\end{array}
\end{equation}
with $\tau = 5$, and 9 Gyr, as indicated. The gas and metallicity
of the stellar population are the same in all models of Table 1,
covering the two cases of $Z_{\ast} = Z_{gas} = Z = 1$ as well as 0.4
\zsun.  The ionization parameter is log~$U = -3.00$. In the electronic
version, the models cover a wider range of the following parameters: $\tau =
1$, 3, 5, 7, and 9 Gyr, and $\infty $ (i. e.  constant SFR), log~$U =
-3.25$, $-$3.00, $-$2.75, $-$2.50 and $-$2.25, $\msup = 40$, 60, 80
and 125 \msun\ and $Z = 0.2$, 0.4, 1.0 and 1.5 \zsun\ (\S 2.2).

In Table 1, column (1) gives the metallicity $Z$, column (2) $\tau$ (see
eq. 1), column (3) \msup\ (in units of \msun), column (4) ${\rm log} U$,
column (5) the age $t$ of the burst (in units of years), column (6) the
hardness of the ionizing MS spectrum $\eta$. In column (7), we list $Q_{H^0}$,
the ionizing photon luminosity of MS stars (s$^{-1}$) per unit solar
mass of the parent galaxy total mass.  column (8) contains the absolute
visual magnitude, $M_V$, per unit solar mass of the parent galaxy
mass.  In columns (9)-(13) we present the equivalent widths of \oii,
\hb, \oiii, \ha, and \nii\ (see
\S 2.4.3).  The emission-line fluxes of \oii, \hb, \oiii, \ha, and
\nii, shown in columns (14)-(18), are in ergs s$^{-1}$ per unit solar
mass of the parent galaxy mass.

\subsection{Relationships between EWs and metallicity $Z$ or age $t$}

The relationship between \wl(\oii) and \wl(\ha) as a function of time
for an exponentially declining SFR (Eq. 1) with a fixed $\tau= 3$ Gyr
is shown in Fig 4.  Each line joins models of increasing age $t$, from
right to left, keeping the ionization parameter constant. The three curves
in each inset correspond to different ionization parameters, covering
the values: log $U$ = $-$2.75, $-$3.00 and $-$3.25 (long-dashed
curve). Models for four different upper mass limits \msup\ of 125, 80,
60 and 40 \msun, are depicted in the four panels of Figs. 4a--d,
respectively.  The model metallicity varies from inset to inset, in a
vertical fashion, to cover the four cases $Z=0.2$ (highest inset),
0.4, 1.0, and 1.5 (lowest inset).  As a reference, a star and a circle
are used to indicate ages of 1 and 12 Gyr, respectively\footnote{The
model ages can be read from Figure 9 (curves labeled $\tau = 3$),
which displays the \ha\ EWs as a function of $t$ (see \S 3.4).}.
Filled squares represent the \joo\ sample (\S 2.3). The same models
are shown in Figures 5a--5d (\msup) for the case of \wl(\oiii)
versus \wl(\ha), while for
\wl(\nii) versus \wl(\oii), they appear in
Figs. 6a--6d. In Figure 7, we plot $\wl(\oiii)+\wl(\oii)$
versus \wl(\ha), assuming the value $\msup\ = 125 \msun$.  We adopt a
logarithmic scale in our figures since \joob\ pointed out that their
EWs plots show a near uniform dispersion when expressed in logarithmic
units. 
 
It is interesting to analyze Figs 4--7 in the light of the discussion
of the data presented in \S 4.2 of \joob. We concluded the
following\footnote{In many
line ratio diagnostic diagrams in the literature,  the \hb\ or \ha\
line appears in the denominator.  In our EW plots involving
\wl(\ha) (Figs. 4, 5, 7 and 8), a normalization is also occurring
implicitly. In the case of line ratios,  the luminosity
of the stellar burst (via $Q_{H^0}$) is taken out when dividing by
\hb. To some extent, this occurs when \wl(\ha) (or any emission-line
equivalent width) is plotted, since it
implies a division of the line luminosity by the underlying continuum
luminosity. A new variable is implicitly introduced in the models,
however, namely the burst age $t$ (or equivalently $\tau$), because
there exist two major contributions to the underlying stellar
continuum: the youngest hot stars, which scale with $Q_{H^0}$, and the
older stellar population, which does not synchronize with the temporal
evolution of the emission lines.}:
\begin{enumerate}

\item Because  high metallicity models appear too far below the data
in Figs 4a--d when $\msup \la 80 \msun$, we consider that our best
case continuous SFR model favors high values for \msup.  For the sake
of the discussion, we will assume $\msup = 125 \msun$ for the
remainder of this Section and the next.
 
\item Although we cannot infer in a simple way the metallicity using
EW diagrams, a comparison of models of different metallicities in
Figure 4a and 5a tells us that galaxies with a high \wl(\oii) and/or a
high \wl(\oiii) (i.e.  objects toward the upper right corner) could
not be fitted satisfactorily using solar or above metallicity
models. Therefore, our models suggest abundances lower than solar for
this subset. This indication is consistent with the findings of \joob,
who pointed out that these high \wl(\oii) objects in fact correspond
to a lower luminosity (and lower metallicity) subset of their
survey\footnote{\joob\ established convincingly that the galaxies with
a \oii\ $vs.$ \hnd\ relation steeper (in linear plots) than the mean
value of 0.4 previously reported by Kennicutt (1992) corresponded in fact
to a lower metallicity galaxy subset if one adopts the calibration of
Zaritsky, Kennicutt, \& Huchra (1994), and Kobulnicky \& Zaritsky
(1999) to infer the galaxy metal abundances.}.

\item The data show a trend of increasing  \wl(\oii) or
\wl(\oiii) with \wl(\ha). Our models suggest that these
could in part be due to a younger age effect, since the solid lines
present an angle of inclination that is comparable to the trend in
the data. However, the increase in
\wl(\oii) or \wl(\oiii) [but not that of \wl(\ha)] may 
be caused as well by a systematic trend in metallicity, since models shift
vertically by large amounts when varying $Z$, as shown by the insets
of Fig 4a.  This is not surprising, given that the luminosities of
collisionally excited lines are strongly dependent on the gas
temperature, which, in turn, is mostly a function of gas metallicity.
We conclude that variations in age $t$ and in metallicity $Z$ are the
two most significant contributors to the trends seen in the data.

\item Even though   a variation in the age $t$ of models  mimics part of
the trend observed in the data in Fig 4a, we cannot reliably estimate 
the age (or the metallicity) using \wl(\oii). For instance,
galaxies in the \joob\ sample with $\wl(\oii) \gtrsim 30 $\AA\
are in general expected to possess subsolar abundances with $Z < 0.5$
\zsun. [In that case, their observed \wl(\ha) appears confined to
values below $\sim 70$ \AA.] However, much larger EW(\oii) than
present in the current sample are in principle possible, but only if
the \ha\ equivalent widths were correspondingly very large. In our
scheme, such an object would correspond to stellar bursts characterized
by a very young stellar population. The predicted \wl(\ha) in that
case can be found amongst the models of Table 1.
 
\item In the case of the \ha\ recombination line, as expected, 
its EW is independent of $U$, while (by definition) high- (low-)
U models are characterized by higher (lower)
\oiii\ EWs and lower (higher) \oii\ EWs. 

\item In models, \wl(\oiii) is much more sensitive to
variations in $U$ (compare Fig. 5a with 4a) than
\wl(\oii). This is consistent with the  range of values taken
by \wl(\oiii) within the data, which are  much larger than that of
\wl(\oii). We   recall that the hardness $\eta$ becomes constant after
an age as short as $\approx 10^7$ \, yr, as shown in \S 2.4.1, and
therefore, changes in \wl(\oiii) are not due to changes in $\eta$, but
are entirely due to changes in $Q_{H^0}$ and in the integrated stellar
continuum level.

\item The \nii--\oii\ EW plot of Fig. 6 indicate that variations in age 
$t$ run almost vertically and are therefore nearly orthogonal with
respect to variations in metallicity $Z$. Furthermore, the models are
not sensitive to the ionization parameter. This EW diagram might
therefore serve as an abundance diagnostic, a property worth
investigating further.

\item The dependence on the ionization parameter  is almost absent in
the plot of $\wl(\oiii)+\wl(\oii)$ against
\wl(\ha), as shown in Fig. 7. 

\end{enumerate}

\subsection{The  abundance--\msup\ degeneracy}

In order to exemplify the degeneracy between abundance and \msup\
(see Sec. 2.4.1) in the case of continuous star formation, we present
in Figure 8 the loci occupied by models with metallicities 0.4, 1.0,
and 1.5 times solar, in a plane defined by (panel a) \wl(\oii) versus
\wl(\ha) and (panel b) \wl (\oiii) versus \wl(\ha).  Areas with
different shades correspond to models with different $Z$-values (as
labeled). The other parameters are log $U$ between $-3.25$, and
$-2.50$, a SFR declining exponentially with time (Eq. 1) with $\tau$
anywhere between 1 Gyr, and $\infty$ (i.e. constant SFR), and a
Salpeter IMF with \msup= 125, 80, and 60 \msun.  These parameters
encompass a wide range of possibilities for the galaxies
considered. The main issue to highlight here is the extent to which
metallicity and \msup\ are degenerate as showed by the significant
overlap between the different shaded areas in the \oiii\ and \ha\ EW
plot of Figure 8b.  The abundance--\msup\ degeneracy\footnote{The
degeneracy in the case of continuous SFR is at least partly
multivariate since the new variable $t$ (via Eq. 1) makes the SED (and
hence the EWs) a function of time due to the evolution of the stellar
continuum (even though the hardness $\eta$ is constant  after
$10^7$ yr).  We
did not explore in detail the boundaries of the $t$--$Z$--\msup\
degeneracy.}  must therefore be taken into account for the analysis of
emission line properties, especially in small \oiii\ equivalent widths
galaxies, in order to determine the origin of the low excitation. Because of the low ionization potential of O to O$^+$, \oii\ does not suffer
from such a degeneracy and it is therefore a better indicator of
abundances in emission-line galaxies.

The comparison of our models with observed line ratios allows us to
conclude that the metallicities of the galaxies in the \joo\ sample
do not extend beyond 1.5 \zsun and that an important fraction of
them actually belongs to a galaxy population of lower metallicity than solar.

\subsection{Varying the rate of SFR decline $\tau$}

Models with different e-folding time $\tau$ (but equal $Z$ and $U$)
show relationships between the \oii\ or the \oiii\ EWs, and the \ha\
EWs, very similar to those studied in \S 3.2, except for their
temporal evolution rate, which scales differently. This is expected since the
hardness of the ionizing SED is constant after $\approx 10^7$ \, yr, as
shown in \S 2.4.1, and models of interest are typically much older than
this. Also, since our EW models correspond to the properties of the
whole host galaxy, the SED in the optical, against which the EW are
defined, is not  sensitive to the youngest star colors.

The equivalent width--age relationship, as expected and mentioned
above, strongly depends on the adopted particular SFR history, as
illustrated by Fig. 9, in which we plot the time evolution of \wl(\ha)
for similar models than shown in previous figures except that $\tau$
is now varied. We only consider   the case $\msup =125 \, \msun$, which
is the   model favored by our discussion in \S 3.2. The horizontal
lines circumscribe the region occupied by the \joo\ galaxies. 
The \wl(\ha) versus age relation is not sensitive to the
ionization parameter. However, it shows a residual dependence on
metallicity, as can be appreciated in Figure 9.  This is simply due to
variations in the mean effective temperature of the ionizing stars
with $Z_{\ast}$, which brings about small changes in the number of
ionizing photons and also to variations in the integrated stellar
continuum.

The first conclusion reached from introspection of Fig. 9 is that even
for the most active galaxies (in the sense of higher star-forming
rates), the stellar population must be older than $\approx$ 1 Gyr in
order to fit within the observed EW range.  Actually, even allowing
for a very long-lasting episode of high SFR, i.e. models with high
$\tau$-values, the requirement that these models lie below the
observed 100 \AA\ \wl(\ha) upper limit, imply such a lower limit on the
burst age, as expected in a sample of normal field galaxies.
On the other hand, we
would conclude that galaxies from other samples, which would be
observed to have \wl(\ha) $> 100$ \AA, must therefore be actively
forming stars, and their dominant stellar population could not be much
older than 1 Gyr.

The second conclusion derived from Fig. 9 is that galaxies found with
a small \wl(\ha) ($\la 10$\AA) are currently forming stars at a very
slow rate with respect to their previous SFR.  This is implied by the
requirement of a small $\tau$ ($\leq$ 3 Gyr) so that  the models
can reproduce \wl(\ha) $\lesssim 10$ \AA.

A more precise determination of the star formation history is 
not possible, using only $W_\lambda$(\ha). As is obvious in Fig. 9,
the same $W_\lambda$(\ha) is obtained by either increasing or decreasing 
$t$ and $\tau$ simultaneously. The degeneration between both parameters
can be overcome by using specific tracers.
Kauffmann et al (2003), for instance, used the strength of the 4000 \AA\
break  as well as the stellar
absorption line index H$\delta$ to constrain the star formation history
of the galaxies. Recently, J. Mateu, G. Bruzual A., \& G. Magris C. 
(2003, in preparation) have model-fitted the 
stellar continuum of the JFFC sample and were able to infer
the  temporal evolution of the SFR of the underlying stellar population
for most of the sample.

As for the line ratios \oii/\hb, \oiii/\hb\ and (\oii$+$\oiii/)\hb,
their behavior as a function of the \ha\ EW is shown in Figure 10 for
models with a Salpeter IMF and $\msup = 125 \, \msun$.  Each strip
represents the loci of models of the same metallicity with log $U$
anywhere between $-3.25$ and $-2.25$ and $\tau$ between 1.0 and
infinity (equivalent to a constant SFR). The corresponding ages of the
models can be inferred from Fig. 9.  Since models with $Z = 0.2$ and
0.4 would overlap, we only show the case $Z=0.4$. 
We infer from panel a that the \oii/\hb\ ratio
could give us a reasonable estimate of abundances,
specially for metallicities above solar,
while the classical abundance indicator
(\oii+\oiii)/\hb\ ratio (Pagel \etal 1979), as expected, remains a
better diagnostic since the shaded strips are much thinner in the
corresponding panel c.
The \oiii/\hb\ ratio in panel b, remains a poor estimator of metal
abundance in the nebulae.

\subsection{General discussion}

Stasi\'nska et al. (2001) presented instantaneous
burst models for the evolution of \hii\ galaxies and suggested that
the presence of an underlying old stellar component was required to
lower the \wl(\hb) predicted by their models sufficiently. Similarly,
Moy et al. (2001) have shown that an instantaneous
or a very short-lasting burst of star formation predicts very high
hydrogen lines equivalent widths. Therefore, they added an underlying
population to their starburst galaxy model. Our integrated continuous
SFR model includes and evolves its own intrinsic older stellar
population in a self-consistent manner. This has the effect of
``raising'' the continuum in the \hb\ and \ha\ wavelength regions,
thereby lowering the nebular emission line equivalent widths.

An important issue to emphasize is that the \ade\ model is able to
simultaneously account for the observed oxygen-to-hydrogen emission-line
	ratios as well as for the \ha\ equivalent widths. In any event,
	the presence of an older population affects EWs but   not  the  line
	ratios\footnote{The line ratios   as a function of age depend
	only on $\eta$, since both $U$ and $Z$ are assumed constant with time.}.
	As has been explained in \S 2.4.1, in the case of a continuous SFR,
	the hardness of the ionizing spectrum evolves only during the first
	few Myr, until a balance between stellar births and deaths is
established. In this latter regime, the parameters that govern the
emission line ratios are the metallicity and the IMF upper mass limit,
and, to a lesser extent, the slope of the initial mass function (a
parameter not explored in this paper).

Even though our approach is not identical to that of Charlot \&
Longhetti (2001), both models have in common that they decouple the
evolution of the ionizing stellar population from the evolution of the
older stellar population since they require us to select, for a fixed
metallicity, the hardness of ionizing spectrum. This is done in our
case by  appropriately  selecting the most suitable IMF upper mass
limit, or, in the case of Charlot \& Longhetti, by selecting the
stellar burst age.

\section{Conclusions}

We have presented a new library of models, \ade, which combine the
stellar population history with the emission line-strengths.  These
models can be used to interpret the properties of objects that are
characterized by either a very recent and strong burst or by a secular
and lower star formation rate.
 
Starting from simple assumptions about the distribution of \hii\
regions in disk galaxies, namely, a uniform distribution of dust-free
gas and stars, in which young stars are distributed in mass in a
similar fashion as the older stars, we have shown how it is possible
to compute several properties of star forming galaxies. A
photoionization code is used to compute the intensity of the emission
lines, assuming no temporal evolution of $U$, while an evolutionary
synthesis code provides the number luminosity and hardness of the MS
ionizing photons. These, in conjunction with the photoionization code
and the integrated stellar distribution, allow us to compute the
temporal evolution of the emission line equivalent widths.

Diagnostic diagrams built by comparing pairs of equivalent widths of
various lines, show that our model predictions fall in figures
in the same region as the observed galaxies of \joo.  In
particular, we conclude the following.
\begin{enumerate}
\item A factor of 10 in $U$ and a range in metallicity between 0.2
and 1.5 \zsun\ seems appropriate to describe the observed properties of
a representative sample of nearby galaxies.
\item The hardness of the ionizing continuum is constant after $10^7$
yr in  case of a continuous SFR.
\item Our favored {\it continuous} SFR model with 
$\msup = 125 \msun$ presents a hardness $\eta$     
very similar to that produced by an instantaneous burst with an IMF
having only $\msup = 60 \msun$.
\item Galaxies with $\wl(\oii)\gtrsim 30$ \AA\ and 
\wl$(\ha) \lesssim 70$ \AA\, appear to have subsolar abundances.
\item Galaxies with $\wl(\ha) \lesssim 100$ \AA\ have a dominant 
underlying stellar 
population older than at least 1 Gyr. Only galaxies with a recent
burst of star formation ($t \lesssim 1$ Gyr) are able to show \ha\ EW
in excess of 100 \AA.
\item The information provided by the \oiii\ line is of limited use,
owing to the strong \msup--$Z$ degeneracy, which complicates the
interpretation of this high-excitation indicator, particularly in the
case of low \oiii\ EWs.
\end{enumerate}

\acknowledgments

We thank J. Mateu, who helped with the installation of the last version
of \map\ at CIDA, and Diethild Starkmeth for proof reading
the paper.  G. M. C. and G. B. A. acknowledge financial support from the
Venenezuelan Ministerio de Ciencia y Tecnolog\'ia a and from FONACIT.
The work of L. B. was supported by the CONACyT (M\'exico) grant 32139-E.

\clearpage

\begin {figure}
\plotone{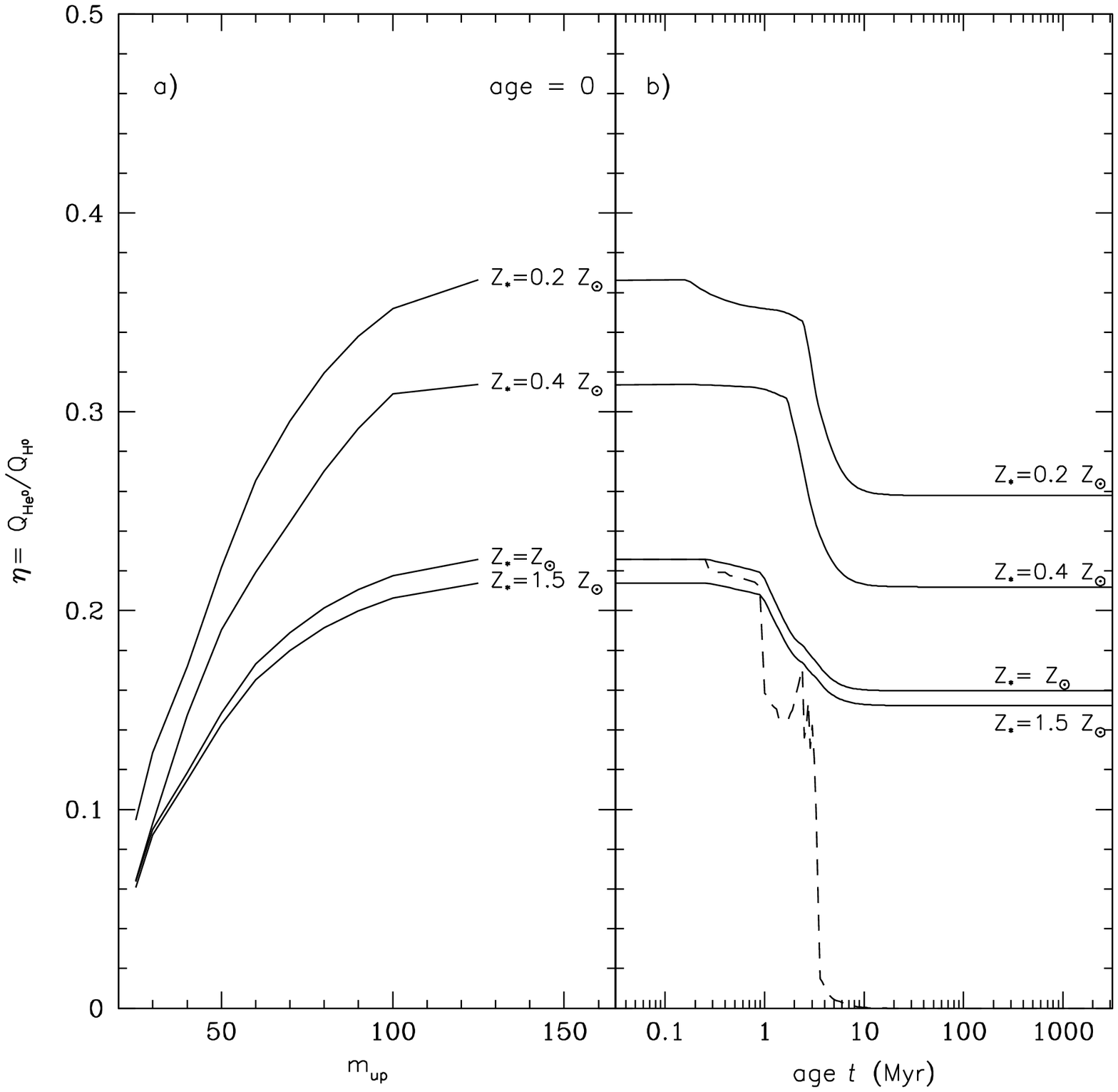}
\caption{{\it (a)} Behavior of $\eta$, the hardness of
the ionizing continuum (as defined in \S 2.4.1), as a function of \msup,
assuming an instantaneous burst with a  Salpeter IMF and one 
of the following metallicities:
$Z_{\ast} = 0.2$, 0.4 , 1.0 and 1.5 \zsun.  {\it (b)} Temporal
evolution of the hardness for the same range in metallicity (as
labeled) in the case of a continuous SFR decreasing exponentially with
time, a Salpeter IMF and the same \msup\ of 125 \msun.   After 10 Myr, $\eta$
reaches an asymptotic value, which differs depending on metallicity, but
can be shown to be independent, of the SFR provided that the SFR is
continuous and either constant or monotonically decreasing. The 
dashed line corresponds to an instantaneous burst of solar metallicity
with the same IMF and \msup.  }
\end{figure}

\begin {figure}
\plotone{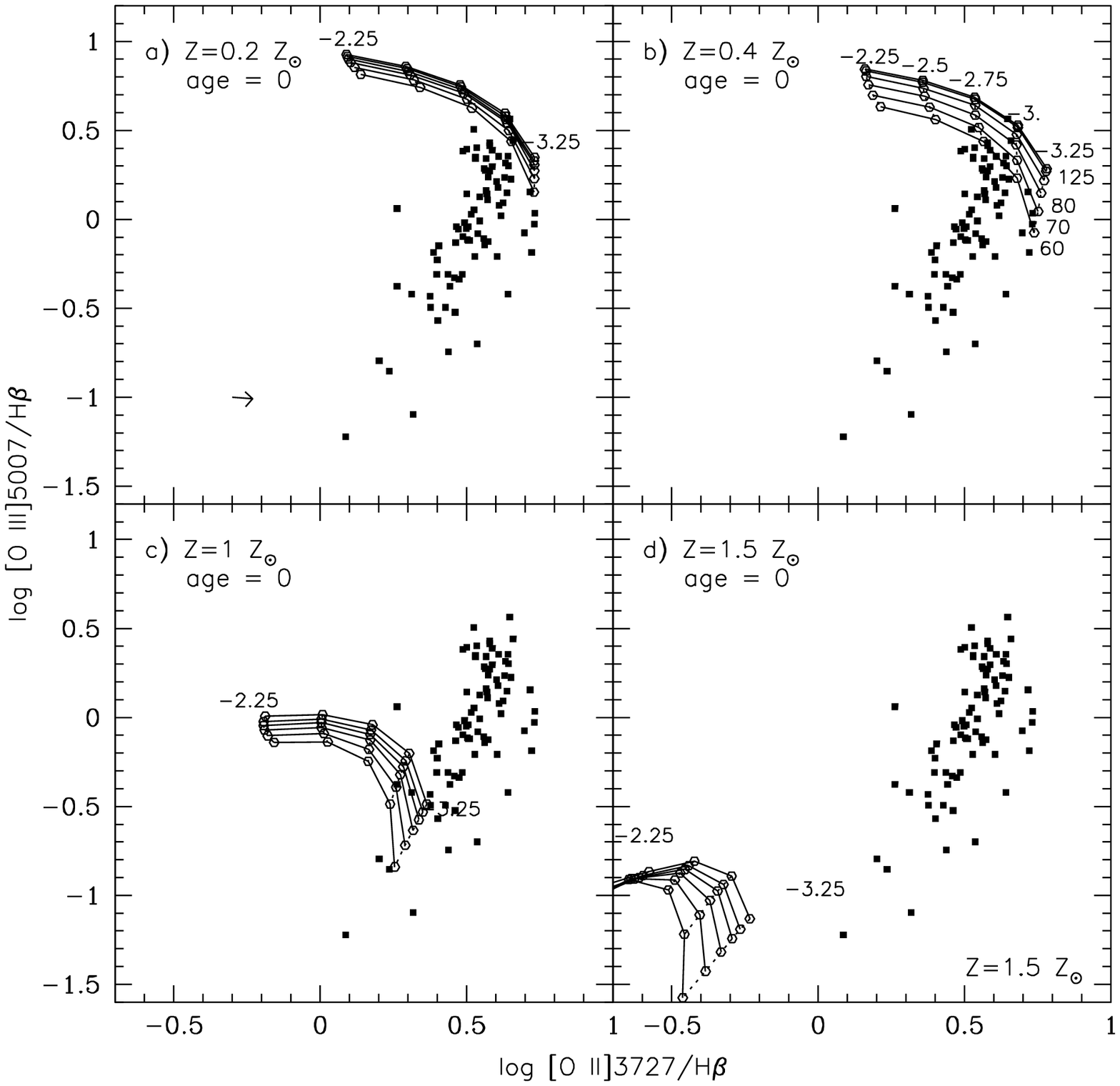}
\caption{Diagnostic plot of log~(\oiii/\hb) 
against log~(\oii/\hb) for {\it (a)} $Z=0.2$ \zsun, {\it (b)} $Z=0.4$
\zsun, {\it (c)} $Z=1.0$ \zsun, and {\it (d)} $Z=1.5$ \zsun.  The
theoretical grid models were calculated with an ionizing spectrum from
an instantaneous burst of age zero, a Salpeter IMF, and $\msup = 125$,
100, 90, 80, 70 and 60 \msun, respectively. The models move up in the
diagram with increasing \msup\ [as labeled in panel (b)].  Solid
lines join models ({\it encircled dots}) of equal \msup,
i.e. constant hardness, while dotted lines join models of equal
ionization parameter $U$. From left to right, log $U$ takes on the
values of $-$2.25, $-$2.50, $-$2.75, $-$3.00, $-$3.25, respectively.
Filled squares represent the observed reddening corrected ratios
from the \joo\ atlas (see \S 2.3). The arrow in {\it (a)} represents the
dereddening vector corresponding to E(B--V) = 0.2.  }
\end{figure}

\begin {figure}
\plotone{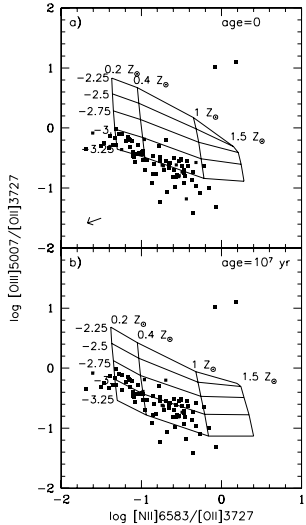}
\caption{
Diagnostic plot of log(\oiii/\oii) against log (\nii/\oii) {\it (a)}:
for an instantaneous burst at zero age, and {\it (b)}: for a
continuous SFR at age 10 Myr, at which point the ionizing hardness $\eta$
has reached its asymptotic value (see Fig. 1b).  In both panels, a
Salpeter IMF with the same $\msup = 125  \msun$ was assumed, and the
nitrogen abundance was increased via secondary nucleosynthesis, as
explained in \S 2.2.  In the grid  shown in each panel,
models are joined by solid lines with the metallicity increasing from
left to right, while the ionization parameter (log $U$) increases going
up.  {\it Filled squares} represent the observed reddening corrected
ratios from the \joo\ atlas. The arrow in {\it (a)} represents the
dereddening vector corresponding to E(B--V) = 0.2.  }
\end{figure}

\begin {figure}
\plotone{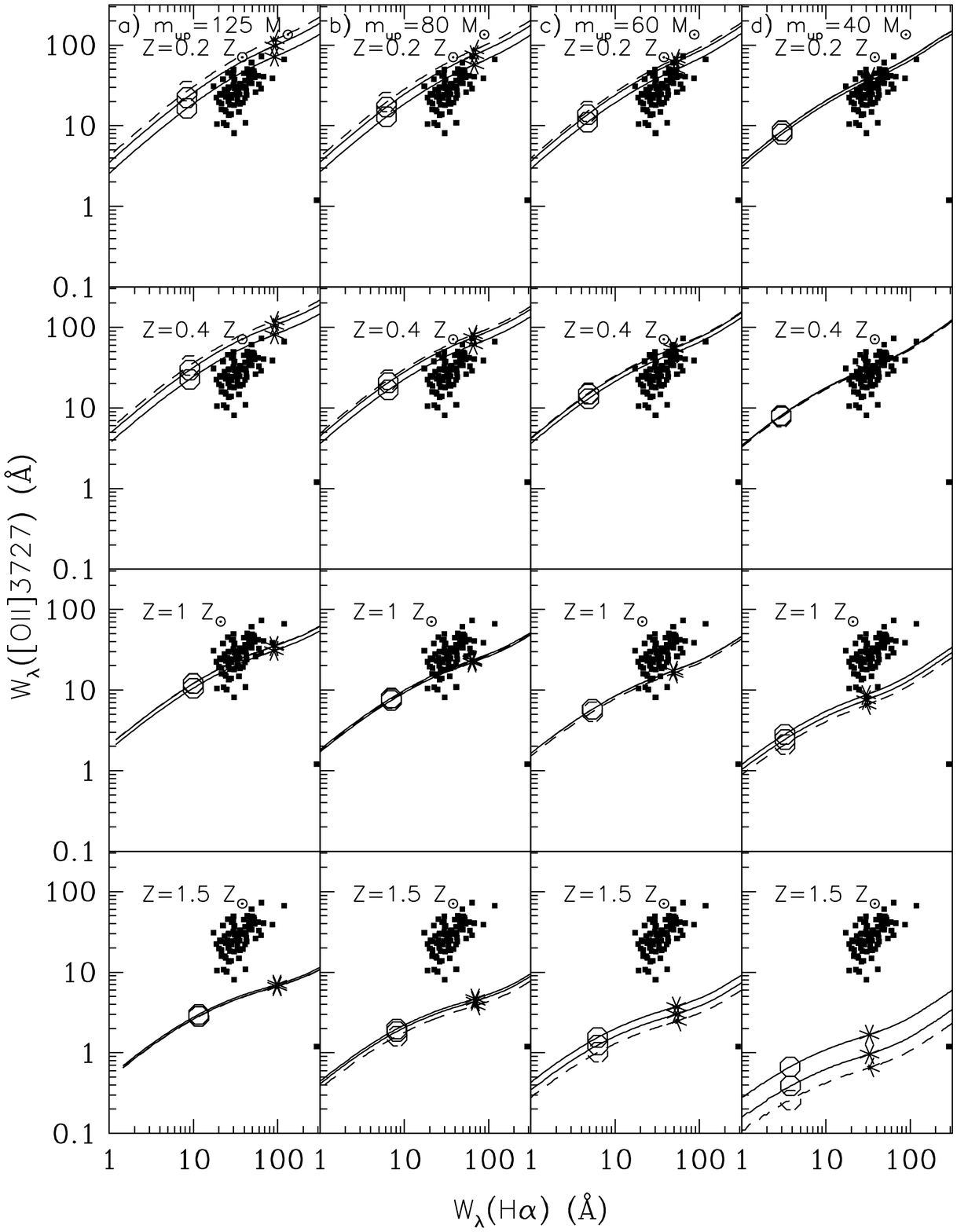}
\caption{
Relation between the equivalent widths of \wl(\oii) and \wl(\ha).
Lines join models of increasing age $t$, from right to left,
that were calculated assuming a continuous exponentially decreasing
SFR (Eq. 1) assuming $\tau = 3$ Gyr. The upper mass limit of the
Salpeter IMF varies from panel to panel as follows: {\it (a)} $\msup =
125 $ \msun, {\it (b)} $\msup = 80$
\msun, {\it (c)} $\msup = 60$ \msun\ and {\it (d)} $\msup = 40$
\msun. Within each panel, different insets correspond to different
metallicities $Z$ of models, increasing from top to bottom: $Z=0.2$
\zsun, $Z=0.4$ \zsun, $Z=1.0$ \zsun, and $Z=1.5$ \zsun.  As a guide, an
asterisk and a circle indicate the position of models with
ages of 1 and 12 Gyr, respectively.  The three lines correspond to a
different ionization parameter each and cover the values log $U$ = $-$2.75,
$-$3.00 and $-$3.25. The long-dashed curve represents the lowest U-value 
model. The filled squares represent reddening
corrected ratios from the
\joo\ atlas, who reported $\pm$3\% to 10\% 
for strong emission lines, and $\pm$30\% for \wl\ smaller than 2 \AA.}
\end{figure}

\begin {figure}
\plotone{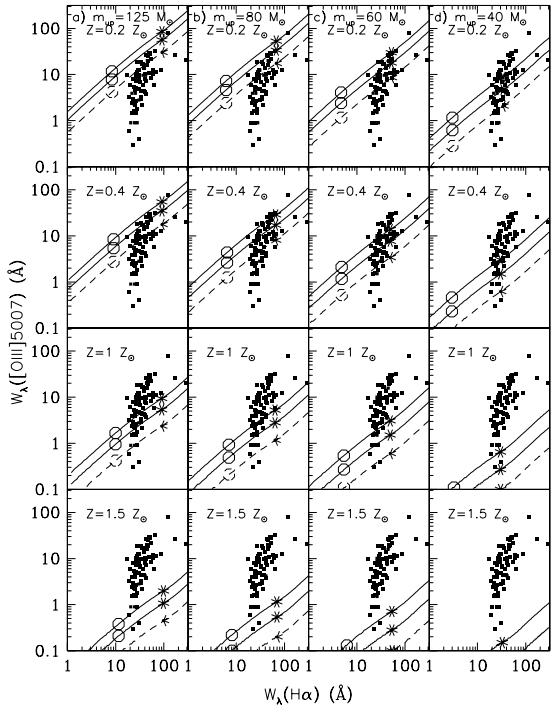}
\caption{
Same as Fig. 4, but for \wl(\oiii\ $\lambda$ 5007) $vs.$ \wl(\ha).  The
long-dashed line represents the lowest U-value model.   }
\end{figure}

\begin {figure}
\plotone{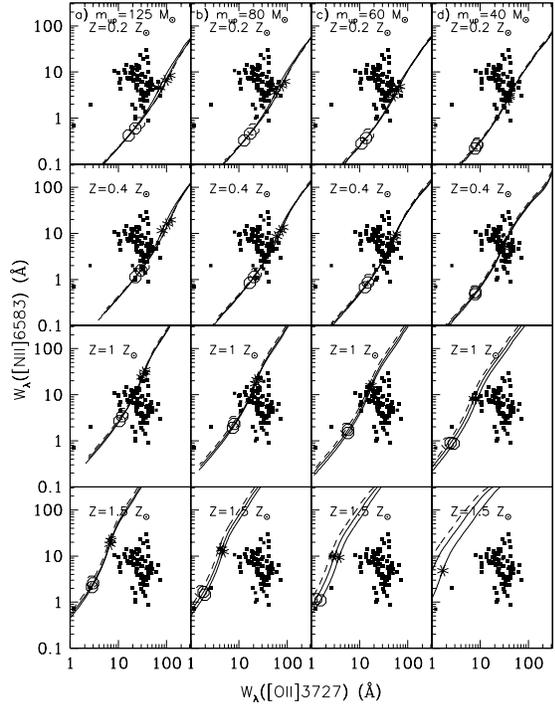}
\caption{
Same as Fig. 4, but for \wl(\nii\ $\lambda$ 6584) $vs.$ \wl(\oii). The
nitrogen abundance was increased via secondary nucleosynthesis, as
explained in \S 2.2.  The long-dashed line represents the lowest U-value
model.  }
\end{figure}

\begin {figure}
\plotone{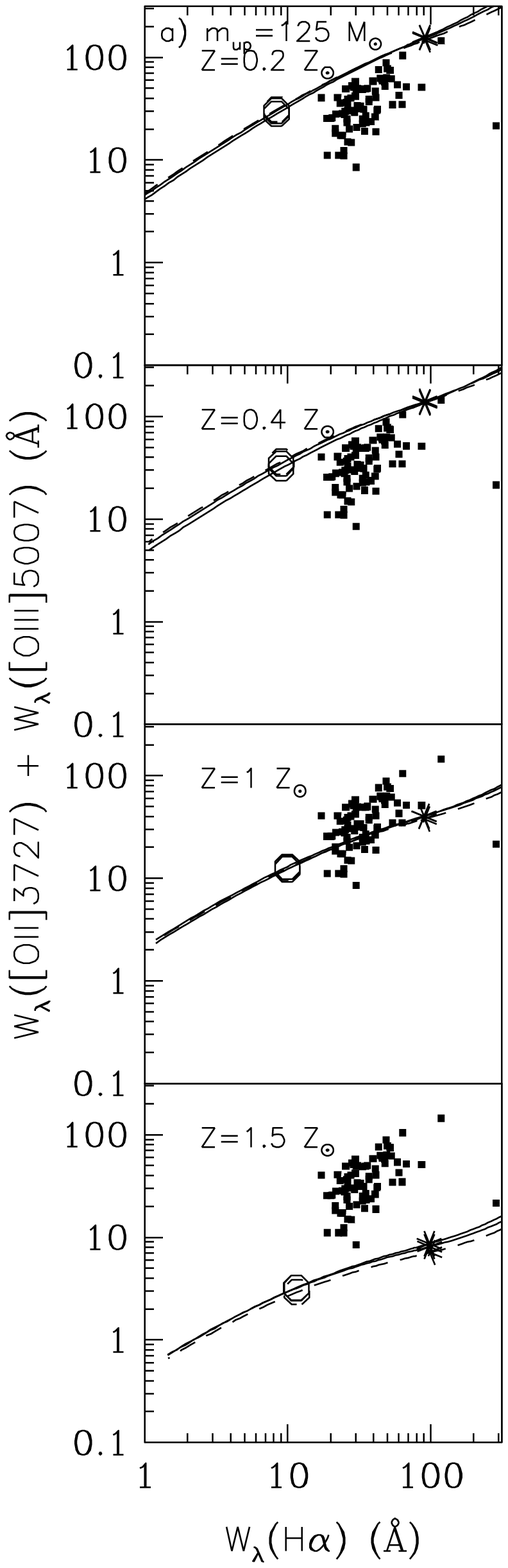}
\caption{
Same as Fig. 4, but for \wl(\oiii)$+$\wl(\oii) $vs.$ \wl(\oii\
$\lambda\lambda$ 3727) assuming a Salpeter IMF with $\msup\ = 125
\msun$. The long-dashed line represents the lowest U-value model.  }
\end{figure}

\begin {figure}
\plotone{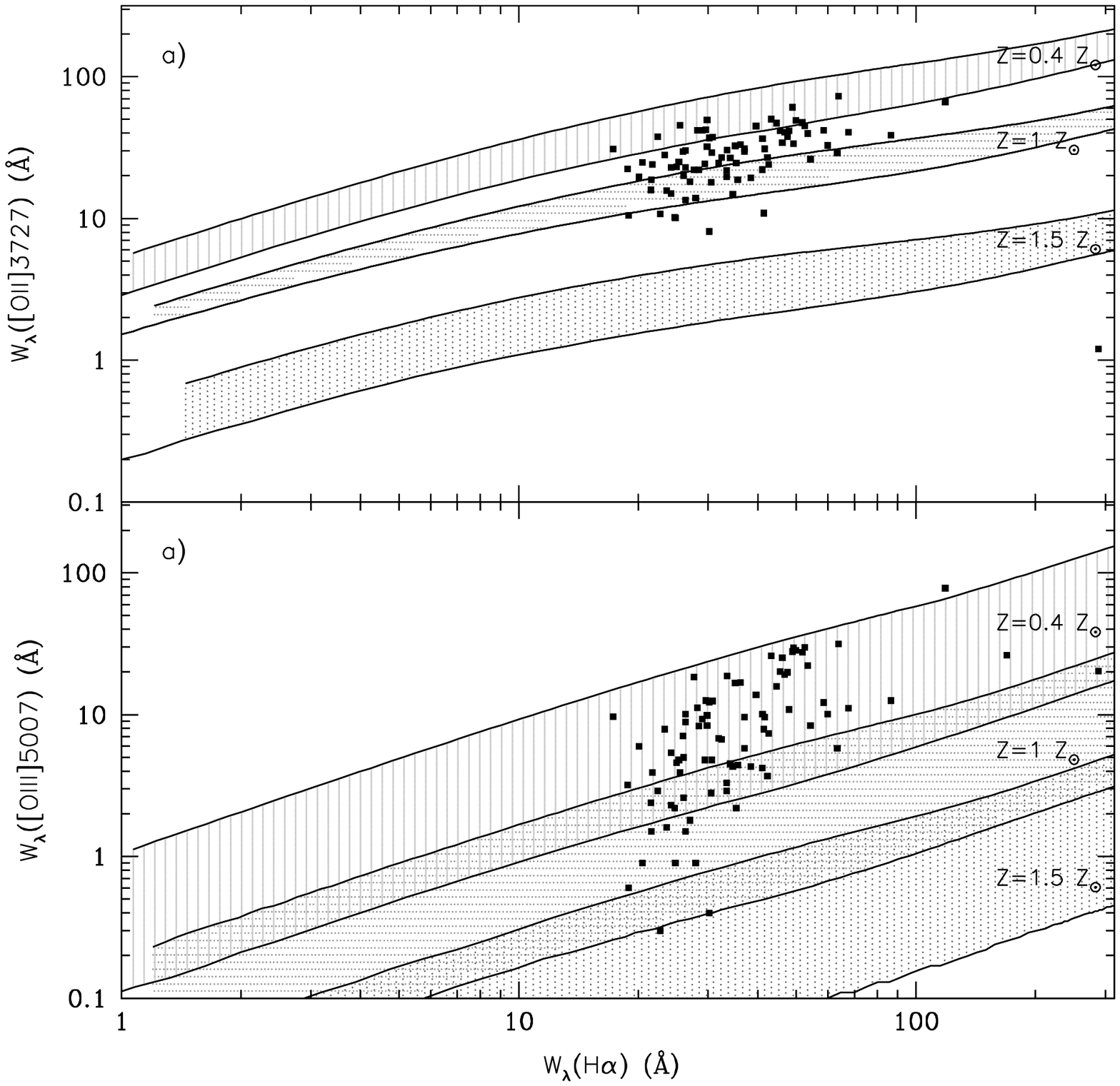}
\caption{
Relation between {\it (a)} \wl(\oii\ $\lambda\lambda$ 3727) $vs.$
\wl(\ha) and
{\it (b)} \wl(\oiii\ $\lambda$ 5007) $vs.$ \wl(\ha). Each shade
corresponds to the loci of continuous SFR models with the same
metallicity (as labeled), but with log $U$ comprised between $-3.25$ 
and $-2.50$, $\tau$ ranging between 1 Gyr, and $\infty$, and
\msup\ taking the values of 125, 80, and 60 \msun. The age $t$ increases
from right to left. The three  metallicities covered are $Z=
0.4$, 1.0 and 1.5 \zsun.  The values observed by \joo\ are represented
in each panel by filled squares.  }
\end{figure}

\begin {figure}
\plotone{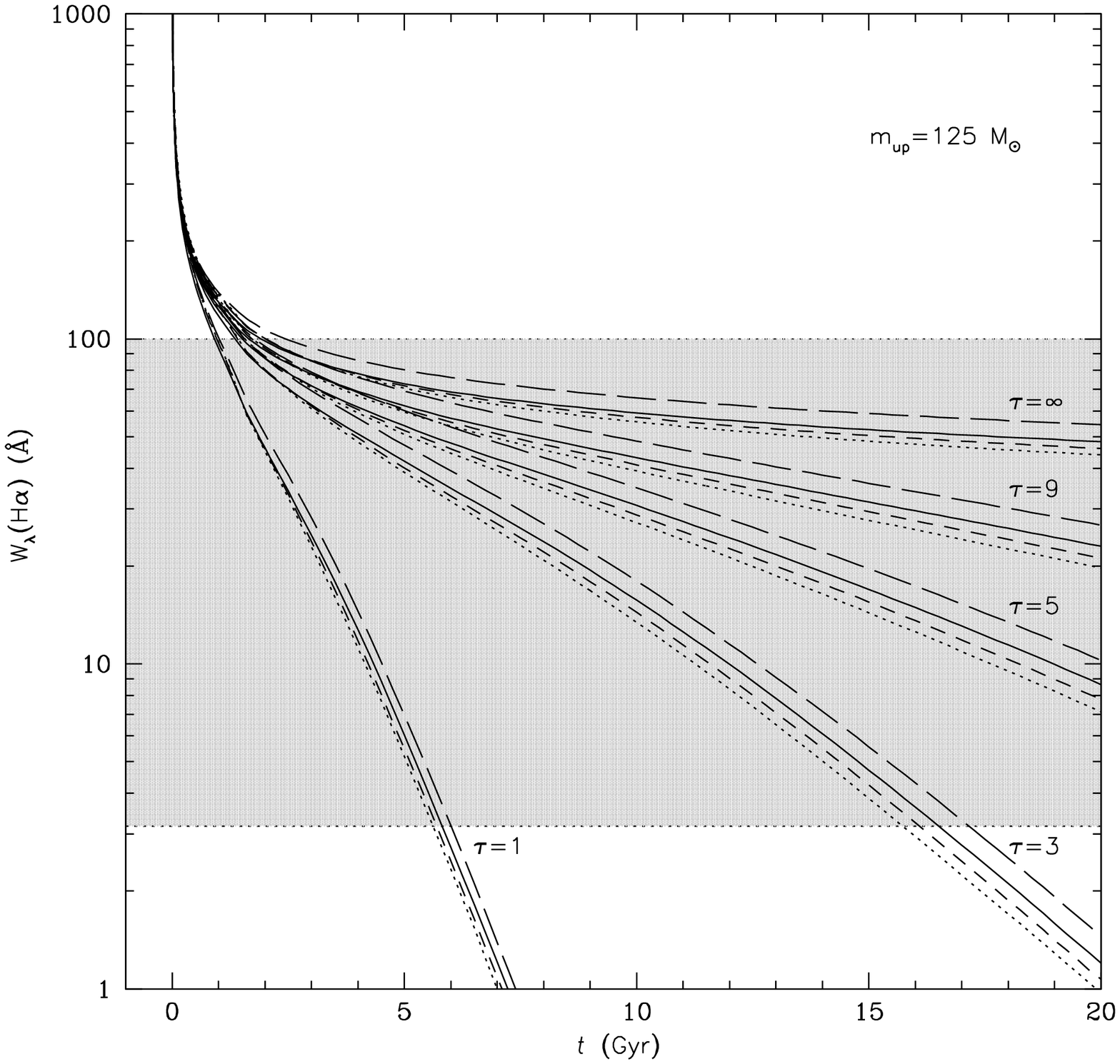}
\caption{
Evolution of \wl(\ha) with age $t$ for models with a SFR declining
exponentially with time, a Salpeter IMF, and $\msup=125 \,
\msun$. The rate of SFR decline is represented by $\tau$, which takes on
the values $\tau$ = 1, 3, 5, and 9 Gyr and $\infty$ (i.e.  constant SFR),
as labeled. The dotted line corresponds to models with $Z=0.2$
\zsun, the short dashed line to $Z=0.4$ \zsun, the solid
line to $Z=1.0$ \zsun, and the long dashed line to $Z=1.5$ \zsun.
The shaded area represents  range of 97\% of observed
\wl(\ha)-values 
from the \joo\ sample of galaxies [three objects exceed the 100 \AA\
bar, having a \wl(\ha) of 169, 288 and 328 \AA].  }
\end{figure}

\begin {figure}
\plotone{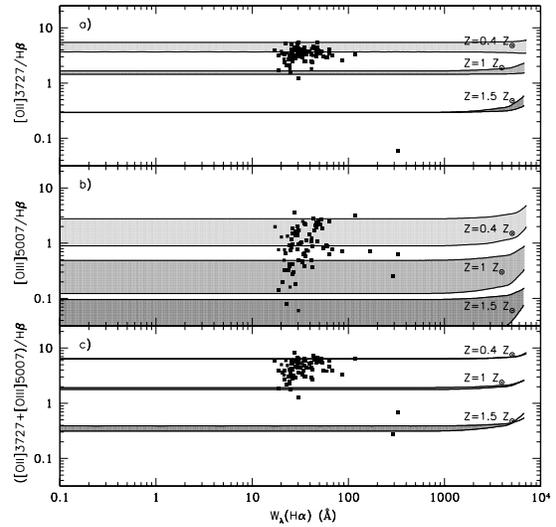}
\caption{
Relation between the line ratios {\it (a)} \oii/\hb, {\it
(b)} \oiii/\hb, {\it (c)} $(\oii + \oiii)/\hb $, and the \ha\
equivalent width.  In each panel, strips with different shades
correspond to models with a different metallicity $Z$, as
labeled. Each shade indicates the loci of models with an exponentially
declining SFR between $\tau = 1$ Gyr and $\infty$. All
models assume a Salpeter IMF and the same $\msup = 125 \msun$.  The
relative thickness of each strip is  mostly due to the dependence on
the ionization parameter, which here spans  the range $-3.25 \le {\rm
log}~U \le -2.50$.  The age $t$ increases from right to left and can
be read from Fig. 9 for any particular value of $\tau$.  Filled
squares represent the \joo\ sample.  }
\end{figure}

\clearpage

\begin{deluxetable}{lcccrrrrrrrrrrr}
\tablecolumns{14}
\tabletypesize{\tiny}
\tablewidth{0pt}
\tablecaption{Model Emission-Line Equivalent Widths and Fluxes}
\tablehead{
\colhead{} & \colhead{}& \colhead{}& \colhead{}& 
\multicolumn{5}{c} {Emission$-$Line Equivalent Widths (\AA)} &
\colhead{} &\multicolumn{5}{c} {log Emission$-$Line Fluxes (erg s$^{-1}$\msun$^{-1}$) } \\
\cline{5-9} \cline{11-15}\\
\colhead{Age} & 
\colhead{$\eta$}&
\colhead{log $Q_{\rm{H^0}}$} &
\colhead{$M_V$\tablenotemark{a}}&
\colhead{[O {\tiny II}]} &
\colhead{H$\beta$} &
\colhead{[O {\tiny III}]} &
\colhead{H$\alpha$} &
\colhead{[N {\tiny II}]} &
&
\colhead{[O {\tiny II}]} &
\colhead{H$\beta$} &
\colhead{[O {\tiny III}]} &
\colhead{H$\alpha$} &
\colhead{[N {\tiny II}]}
\\       
\colhead{(Gyr)} & 
\colhead{} &
\colhead{s$^{-1}$\msun$^{-1}$} &
\colhead{} &
\colhead{$\lambda 3727$} &
\colhead{} &
\colhead{$\lambda 5007$} &
\colhead{} &
\colhead{$\lambda 6583$} &
&
\colhead{$\lambda 3727$} &
\colhead{}& 
\colhead{$\lambda 5007$} &
\colhead{}& 
\colhead{$\lambda 6583$} 
\\       
\colhead{}& 
\colhead{}& 
\colhead{}& 
\colhead{}& 
\colhead{}& 
\colhead{}& 
\colhead{}& 
\colhead{}& 
\colhead{}&
& 
\colhead{}& 
\colhead{}& 
\colhead{}& 
\colhead{}& 
\colhead{} 
}
\startdata
\cutinhead{ $\tau = 5\  \rm{ Gyr}   \ \ \ \ \ \ Z = 0.4 $\zsun \ \ \ \ \ \ log~$U = -3.00$}
  0.5 $\times 10^{-3}$ &    0.31 &   42.49 &   11.45 & 1713.96 &  812.99 & 3030.13 & 7049.33 & 1229.11& &   30.81 &   30.13 &   30.65 &   30.58 &   29.83 \\
  1.0 $\times 10^{-3}$ &    0.31 &   42.79 &   10.70 & 1707.53 &  810.10 & 2987.53 & 7024.48 & 1224.76& &   31.11 &   30.42 &   30.94 &   30.88 &   30.13 \\
  0.5 $\times 10^{-2}$ &    0.22 &   43.32 &    8.46 &  783.72 &  352.06 &  757.18 & 2960.98 &  516.38& &   31.63 &   30.95 &   31.24 &   31.41 &   30.66 \\
  0.1 $\times 10^{-1}$ &    0.21 &   43.34 &    7.54 &  468.70 &  171.76 &  332.61 & 1154.30 &  201.92& &   31.66 &   30.98 &   31.22 &   31.43 &   30.68 \\
  0.5 $\times 10^{-1}$ &    0.21 &   43.34 &    6.67 &  250.54 &   78.35 &  149.24 &  486.97 &   85.22& &   31.65 &   30.98 &   31.22 &   31.43 &   30.68 \\
  0.1 $\times 10^{ 0}$ &    0.21 &   43.33 &    6.36 &  206.15 &   58.68 &  111.53 &  358.92 &   62.81& &   31.65 &   30.97 &   31.21 &   31.43 &   30.67 \\
  0.5 $\times 10^{ 0}$ &    0.21 &   43.30 &    5.70 &  139.53 &   30.67 &   57.67 &  173.08 &   30.29& &   31.61 &   30.94 &   31.18 &   31.39 &   30.64 \\
  1.0 $\times 10^{ 0}$ &    0.21 &   43.25 &    5.51 &  121.50 &   23.84 &   44.53 &  126.85 &   22.20& &   31.57 &   30.89 &   31.13 &   31.35 &   30.59 \\
  2.0 $\times 10^{ 0}$ &    0.21 &   43.17 &    5.41 &  105.32 &   18.54 &   34.40 &   88.27 &   15.45& &   31.49 &   30.81 &   31.05 &   31.26 &   30.51 \\
  4.0 $\times 10^{ 0}$ &    0.21 &   42.99 &    5.51 &   90.32 &   14.28 &   26.34 &   60.50 &   10.59& &   31.31 &   30.63 &   30.88 &   31.09 &   30.33 \\
  6.0 $\times 10^{ 0}$ &    0.21 &   42.82 &    5.70 &   80.11 &   11.70 &   21.50 &   45.93 &    8.04& &   31.14 &   30.46 &   30.70 &   30.92 &   30.16 \\
  8.0 $\times 10^{ 0}$ &    0.21 &   42.65 &    5.91 &   71.70 &    9.79 &   17.95 &   36.28 &    6.35& &   30.96 &   30.28 &   30.53 &   30.74 &   29.99 \\
  1.0 $\times 10^{ 1}$ &    0.21 &   42.47 &    6.13 &   63.81 &    8.16 &   14.93 &   28.74 &    5.03& &   30.79 &   30.11 &   30.35 &   30.57 &   29.81 \\
  1.2 $\times 10^{ 1}$ &    0.21 &   42.30 &    6.33 &   56.17 &    6.73 &   12.29 &   22.61 &    3.96& &   30.62 &   29.94 &   30.18 &   30.40 &   29.64 \\
  1.4 $\times 10^{ 1}$ &    0.21 &   42.12 &    6.52 &   48.79 &    5.47 &    9.98 &   17.63 &    3.09& &   30.44 &   29.76 &   30.01 &   30.22 &   29.47 \\
\cutinhead{$\tau = 5\ \rm{ Gyr} \ \ \ \ \ \ Z =  $\zsun \ \ \ \ \ \ log~$U = -3.00$}
  0.5 $\times 10^{-3}$ &    0.22 &   42.51 &   11.29 &  632.49 &  722.13 &  484.56 & 6671.10 & 2513.29& &   30.44 &   30.14 &   29.92 &   30.62 &   30.19 \\
  1.0 $\times 10^{-3}$ &    0.22 &   42.80 &   10.51 &  596.35 &  696.85 &  436.64 & 6430.04 & 2398.27& &   30.73 &   30.44 &   30.19 &   30.91 &   30.48 \\
  0.5 $\times 10^{-2}$ &    0.17 &   43.25 &    8.21 &  200.84 &  243.36 &   82.78 & 2139.10 &  741.40& &   31.12 &   30.89 &   30.38 &   31.37 &   30.91 \\
  0.1 $\times 10^{-1}$ &    0.16 &   43.26 &    7.49 &  123.64 &  136.00 &   42.22 &  988.10 &  339.00& &   31.12 &   30.91 &   30.35 &   31.38 &   30.92 \\
  0.5 $\times 10^{-1}$ &    0.16 &   43.26 &    6.72 &   74.85 &   69.52 &   21.21 &  437.01 &  149.74& &   31.12 &   30.91 &   30.35 &   31.38 &   30.91 \\
  0.1 $\times 10^{ 0}$ &    0.16 &   43.26 &    6.42 &   62.55 &   52.77 &   16.05 &  324.48 &  111.18& &   31.12 &   30.90 &   30.34 &   31.37 &   30.91 \\
  0.5 $\times 10^{ 0}$ &    0.16 &   43.22 &    5.80 &   44.37 &   28.30 &    8.53 &  164.38 &   56.32& &   31.08 &   30.86 &   30.31 &   31.34 &   30.87 \\
  1.0 $\times 10^{ 0}$ &    0.16 &   43.18 &    5.63 &   39.32 &   22.22 &    6.66 &  121.26 &   41.55& &   31.04 &   30.82 &   30.26 &   31.30 &   30.83 \\
  2.0 $\times 10^{ 0}$ &    0.16 &   43.09 &    5.55 &   34.85 &   17.70 &    5.28 &   87.03 &   29.82& &   30.95 &   30.74 &   30.18 &   31.21 &   30.74 \\
  4.0 $\times 10^{ 0}$ &    0.16 &   42.92 &    5.69 &   30.83 &   14.04 &    4.17 &   61.56 &   21.09& &   30.78 &   30.56 &   30.00 &   31.03 &   30.57 \\
  6.0 $\times 10^{ 0}$ &    0.16 &   42.75 &    5.90 &   28.06 &   11.76 &    3.48 &   47.67 &   16.33& &   30.60 &   30.39 &   29.83 &   30.86 &   30.40 \\
  8.0 $\times 10^{ 0}$ &    0.16 &   42.57 &    6.13 &   25.66 &   10.01 &    2.96 &   38.23 &   13.10& &   30.43 &   30.21 &   29.66 &   30.69 &   30.22 \\
  1.0 $\times 10^{ 1}$ &    0.16 &   42.40 &    6.36 &   23.35 &    8.47 &    2.50 &   30.64 &   10.50& &   30.26 &   30.04 &   29.48 &   30.51 &   30.05 \\
  1.2 $\times 10^{ 1}$ &    0.16 &   42.22 &    6.58 &   21.08 &    7.09 &    2.09 &   24.38 &    8.35& &   30.08 &   29.87 &   29.31 &   30.34 &   29.88 \\
  1.4 $\times 10^{ 1}$ &    0.16 &   42.05 &    6.78 &   18.81 &    5.85 &    1.73 &   19.16 &    6.57& &   29.91 &   29.69 &   29.14 &   30.17 &   29.70 \\
\cutinhead{$\tau = 9\ \rm{ Gyr} \ \ \ \ \ \ Z = 0.4 $\zsun \ \ \ \ \ \ log~$U = -3.00$}
  0.5 $\times 10^{-3}$ &    0.31 &   42.23 &   12.09 & 1713.96 &  812.99 & 3030.13 & 7049.33 & 1229.11& &   30.55 &   29.87 &   30.40 &   30.33 &   29.57 \\
  1.0 $\times 10^{-3}$ &    0.31 &   42.53 &   11.33 & 1707.53 &  810.10 & 2987.53 & 7024.48 & 1224.76& &   30.85 &   30.17 &   30.69 &   30.63 &   29.87 \\
  0.5 $\times 10^{-2}$ &    0.22 &   43.06 &    9.09 &  783.78 &  352.10 &  757.31 & 2961.28 &  516.43& &   31.38 &   30.70 &   30.99 &   31.16 &   30.40 \\
  0.1 $\times 10^{-1}$ &    0.21 &   43.08 &    8.18 &  468.82 &  171.81 &  332.74 & 1154.68 &  201.99& &   31.40 &   30.72 &   30.97 &   31.18 &   30.42 \\
  0.5 $\times 10^{-1}$ &    0.21 &   43.08 &    7.30 &  250.82 &   78.45 &  149.46 &  487.69 &   85.35& &   31.40 &   30.72 &   30.97 &   31.18 &   30.42 \\
  0.1 $\times 10^{ 0}$ &    0.21 &   43.08 &    6.99 &  206.58 &   58.84 &  111.84 &  359.93 &   62.99& &   31.40 &   30.72 &   30.96 &   31.18 &   30.42 \\
  0.5 $\times 10^{ 0}$ &    0.21 &   43.06 &    6.31 &  140.73 &   31.06 &   58.42 &  175.48 &   30.71& &   31.38 &   30.70 &   30.94 &   31.16 &   30.40 \\
  1.0 $\times 10^{ 0}$ &    0.21 &   43.04 &    6.08 &  123.42 &   24.42 &   45.64 &  130.39 &   22.82& &   31.35 &   30.68 &   30.92 &   31.13 &   30.38 \\
  2.0 $\times 10^{ 0}$ &    0.21 &   42.99 &    5.91 &  108.53 &   19.43 &   36.08 &   93.52 &   16.37& &   31.31 &   30.63 &   30.87 &   31.09 &   30.33 \\
  4.0 $\times 10^{ 0}$ &    0.21 &   42.89 &    5.87 &   95.89 &   15.66 &   28.94 &   68.09 &   11.92& &   31.21 &   30.53 &   30.77 &   30.99 &   30.23 \\
  6.0 $\times 10^{ 0}$ &    0.21 &   42.79 &    5.93 &   88.22 &   13.56 &   24.97 &   55.40 &    9.69& &   31.11 &   30.43 &   30.68 &   30.89 &   30.14 \\
  8.0 $\times 10^{ 0}$ &    0.21 &   42.70 &    6.03 &   82.50 &   12.09 &   22.23 &   47.26 &    8.27& &   31.02 &   30.34 &   30.58 &   30.80 &   30.04 \\
  1.0 $\times 10^{ 1}$ &    0.21 &   42.60 &    6.14 &   77.55 &   10.89 &   19.99 &   40.99 &    7.17& &   30.92 &   30.24 &   30.49 &   30.70 &   29.94 \\
  1.2 $\times 10^{ 1}$ &    0.21 &   42.51 &    6.26 &   73.02 &    9.84 &   18.04 &   35.79 &    6.26& &   30.82 &   30.15 &   30.39 &   30.60 &   29.85 \\
  1.4 $\times 10^{ 1}$ &    0.21 &   42.41 &    6.38 &   68.77 &    8.90 &   16.31 &   31.39 &    5.49& &   30.73 &   30.05 &   30.29 &   30.51 &   29.75 \\
\cutinhead{$\tau = 9\ \rm{ Gyr} \ \ \ \ \ \ Z = $\zsun \ \ \ \ \ \ log~$U = -3.00$}
  0.5 $\times 10^{-3}$ &    0.22 &   42.25 &   11.92 &  632.49 &  722.13 &  484.56 & 6671.10 & 2513.29& &   30.19 &   29.89 &   29.67 &   30.36 &   29.94 \\
  1.0 $\times 10^{-3}$ &    0.22 &   42.54 &   11.15 &  596.35 &  696.86 &  436.65 & 6430.05 & 2398.27& &   30.47 &   30.18 &   29.93 &   30.66 &   30.23 \\
  0.5 $\times 10^{-2}$ &    0.17 &   43.00 &    8.85 &  200.86 &  243.39 &   82.80 & 2139.35 &  741.49& &   30.86 &   30.64 &   30.12 &   31.11 &   30.65 \\
  0.1 $\times 10^{-1}$ &    0.16 &   43.01 &    8.13 &  123.67 &  136.03 &   42.23 &  988.42 &  339.12& &   30.87 &   30.65 &   30.10 &   31.13 &   30.66 \\
  0.5 $\times 10^{-1}$ &    0.16 &   43.01 &    7.36 &   74.92 &   69.60 &   21.24 &  437.63 &  149.95& &   30.87 &   30.65 &   30.09 &   31.12 &   30.66 \\
  0.1 $\times 10^{ 0}$ &    0.16 &   43.01 &    7.05 &   62.67 &   52.91 &   16.09 &  325.37 &  111.48& &   30.86 &   30.65 &   30.09 &   31.12 &   30.66 \\
  0.5 $\times 10^{ 0}$ &    0.16 &   42.99 &    6.41 &   44.72 &   28.65 &    8.64 &  166.57 &   57.07& &   30.84 &   30.63 &   30.07 &   31.10 &   30.64 \\
  1.0 $\times 10^{ 0}$ &    0.16 &   42.96 &    6.19 &   39.88 &   22.75 &    6.82 &  124.56 &   42.68& &   30.82 &   30.60 &   30.05 &   31.08 &   30.61 \\
  2.0 $\times 10^{ 0}$ &    0.16 &   42.92 &    6.05 &   35.79 &   18.50 &    5.52 &   91.97 &   31.51& &   30.77 &   30.56 &   30.00 &   31.03 &   30.57 \\
  4.0 $\times 10^{ 0}$ &    0.16 &   42.82 &    6.04 &   32.43 &   15.29 &    4.55 &   68.83 &   23.58& &   30.68 &   30.46 &   29.91 &   30.93 &   30.47 \\
  6.0 $\times 10^{ 0}$ &    0.16 &   42.72 &    6.12 &   30.40 &   13.47 &    4.00 &   56.93 &   19.51& &   30.58 &   30.36 &   29.81 &   30.84 &   30.37 \\
  8.0 $\times 10^{ 0}$ &    0.16 &   42.63 &    6.23 &   28.84 &   12.16 &    3.60 &   49.16 &   16.85& &   30.48 &   30.27 &   29.71 &   30.74 &   30.28 \\
  1.0 $\times 10^{ 1}$ &    0.16 &   42.53 &    6.35 &   27.47 &   11.07 &    3.28 &   43.02 &   14.74& &   30.39 &   30.17 &   29.61 &   30.65 &   30.18 \\
  1.2 $\times 10^{ 1}$ &    0.16 &   42.43 &    6.48 &   26.22 &   10.10 &    2.99 &   37.89 &   12.98& &   30.29 &   30.08 &   29.52 &   30.55 &   30.08 \\
  1.4 $\times 10^{ 1}$ &    0.16 &   42.34 &    6.61 &   25.04 &    9.22 &    2.73 &   33.45 &   11.46& &   30.19 &   29.98 &   29.42 &   30.45 &   29.99 \\

\enddata

\tablenotetext{a}{in mag per galaxy mass in units of solar mass}
\tablecomments{The complete version of this table is in the electronic edition of
the Journal.  The printed edition contains only a sample}
\end{deluxetable}		     	     

\end{document}